\documentclass[12pt]{article}
\usepackage{hyperref}
\usepackage{graphicx}
\newcommand{\figscale}{0.49}
\newcommand{\altfigscale}{0.5}
\date{19 September, 2003}

\title{The ST7 Interferometer}
\author{Robert Spero\qquad Andreas Kuhnert\\ Jet Propulsion Laboratory,
California Institute of Technology}
\begin{document}

\maketitle
\begin{abstract}
Two homodyne Michelson interferferometers aboard the LISA Pathfinder
spacecraft will measure the the positions of two free-floating test
masses, as part of the NASA ST7 mission.  The interferometer is required to
measure the separation between the test masses with
sensitivity of 30\,pm/$\sqrt{\rm Hz}$ at 10\,mHz.  The readout scheme is
described, error sources are analyzed, and experimental results are
presented.
\end{abstract}
\section{Introduction and Interferometer Description}
The ST7 project is designed to demonstrate that test masses can 
be contained within a spacecraft, and left free from external
disturbances at a level approximately 10 times higher than the nominal
requirement for LISA~\cite{PPA2} .  The ST7 electrostatic system that contains,
senses, and controls the test masses is designed to apply forces no
larger than those that would lead to this displacement noise level, and
the interferometer is designed to verify the absence of imposed force
noise.

\begin{figure}
\begin{center}
\includegraphics[scale=\altfigscale]{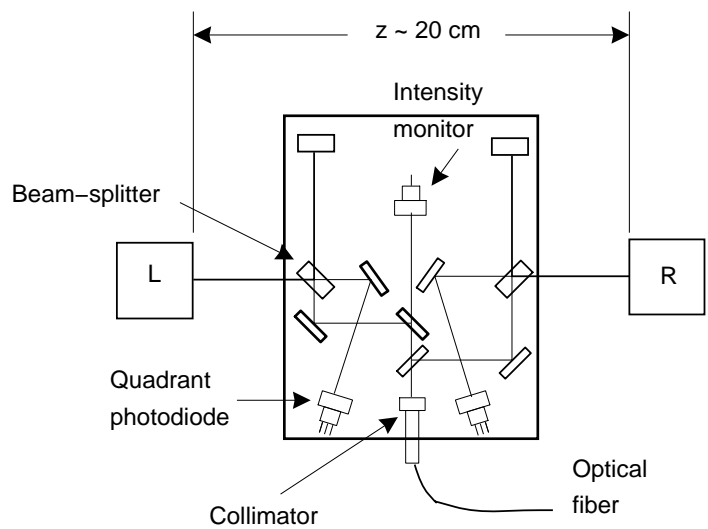}
\caption{Schematic diagram of the ST7 interferometer.  The distances to
the L and R test masses are read out separately, and their separation is
computed as the sum.  The ``sensitive path'' bold lines indicate
the distances measured by the interferometer.
\label{diag}}
\end{center}
\end{figure}

Dual homodyne Michelson interferometers (Figure~\ref{diag}) sense changes in
the positions of the test masses.  Centered 
on a common optical bench, there is one interferometer for
each mass.  The quantity of
interest is $z$, the separation of test masses, which is
equal to the sum of the two interferometer displacement  outputs.  The
signal $I_s$  is the total detected intensity for one
interferometer; at the nominal operating point, $I_s$ is  half its
maximum.
Changes in position of one of the test masses are
detected as changes in  the corresponding $I_s.$

Fluctuation in the orientation $\theta$ of the test mass is a source of
error in the displacement readout.  The signal $I_d$ is the difference
in intensity between two halves of the beam, as monitored by segmented
photodiodes.  $I_d$ measures $\theta$; it can be used to reduce the
error either during data analysis, or  in real-time by serving as  the
error-point in a control system.  

As depicted in
Figure~\ref{diag},  light a 1.06\,$\mu$m wavelength NPRO laser 
enters on an optical fiber,  and is collimated and converted
to a free-space beam of approximately 1\,mm diameter.  The beam is split to
feed two separate interferometers.  
The sum of the reference lengths (drawn vertically)
is matched to the sum of the measurement lengths
(drawn horizontally), to make the measurement of $z$
insensitive to laser frequency.  During data analysis, $I_s$ and $I_d$
are divided by the
laser Intensity Monitor signal picked off before the interferometer, to 
form signals independent of the laser intensity.  

This homodyne detection
scheme has the advantage of simplicity:  laser light is injected by a single
fiber, there are no polarizers or modulators, and the displacement
measurement is derived directly from the detected intensity.  It
requires that the masses be constrained to near the mid-fringe
position, and that the detection electronics have low noise in the
signal band of 1\,mHz to 1\,Hz.  Calculations and measurements indicate
that noise in the laser intensity and frequency and in the readout
electronics are all sufficiently small to meet the ST7 performance requirements.
A potential source of significant noise is fluctuation in $\theta$ 
giving a false reading of noise in $z.$
We emphasize in this paper the mitigation of test mass
orientation noise as sensed by the quadrant photodetectors.

\section{Detected Signals and Angle Sensitivity}
\subsection{Total intensity}
The intensity profile across the interfering beams at the 
photodetectors is determined by
the addition of electric fields, $E_1$ from the reference mirror and
$E_2$ from the test mass.  The beams are like-polarized, and each has a
Gaussian profile with radius $w.$  The optical phase $\phi$ of $E_2$ 
is proportional to the position of the test mass.    If $E_2$ is
misaligned with respect to $E_1$ in the $x$-direction by the slight
angle $\theta,$
\begin{eqnarray}
E_1&=&e^{-r^2/w^2}\\
E_2&=&e^{i(\phi+kx\sin\theta) -r^2/w^2},
\end{eqnarray}
where the wavenumber $k$ is related to the wavelength $\lambda$ by
$k=2\pi/\lambda.$
The resulting intensity at the plane of detection is
\begin{equation}
\label{Irpsi}
I(r,\psi)=|E_1+E_2|^2=2e^{-2r^2/w^2}\left[1+
\cos(\phi+kx\sin\theta )\right],
\end{equation}
where $(r, \psi)$ are the polar coordinates of the intensity
profile and $x=r\sin\psi$.  

Integrating Equation~\ref{Irpsi} over all $(r, \psi$) and normalizing to
unit maximum,   the total 
intensity of the interference signal is
\begin{equation}
\label{intens-sum}
I_s=\frac{1+\cos\phi
\exp(-k^2w^2\theta^2/8)} {2}.
\end{equation}
Equation~\ref{intens-sum} identifies the Visibility, $V,$ defined as
(max - min)/(max + min) of $I_s$, as 
$V(\theta)=\exp(-k^2w^2\theta^2/8),$
and
\begin{equation}
\label{Isvis}
I_s=\frac 1 2 \left(1+V\cos\phi\right).
\end{equation}

The sensitivity of $I_s$ to alignment change is 
\begin{equation}
\label{intens-sum-deriv}
\frac{dI_s}{d\theta}
= -\frac 1 8 k^2 w^2 \theta V \cos \phi 
\approx-\frac 1 8 k^2 w^2 \theta \cos \phi
\left(1-\frac 1 8 (kw\theta)^2\right),
\end{equation}
where the expansion is valid for small misalignment $\theta.$

\subsection{Difference signal}
Consider the division of the beam into upper $(0<\psi<\pi)$ and lower
$(-\pi<\psi<0)$ halves:
\begin{equation}
	I_{u,l}=\int_0^{\pi,-\pi}d\psi
	\int_0^{\infty}rdr\,I(r,\psi),
	=-\frac 1 4 \left[1+V\left(\cos\phi \pm \sin\phi\,\mathrm {erfi}
(\sqrt{-\log V})\right)\right]
\end{equation}
where erfi is the complex error function.
Using the identity
$\cos\phi + A\sin\phi = \cos(\phi-\phi_0)/\cos\phi_0,$
with $\phi_0=\arctan A,$ 
the phase shift between $I_u$ and $I_l$ is
\begin{equation}
\label{delta-phi}
\Delta\phi = 2\arctan\left[\mathrm {erfi}\sqrt{-\log V}\right].
\end{equation}
$V$ can be maximized by aligning the interferometer to minimize
$\Delta\phi$.

The magnitude of the difference signal, $I_d=I_u-I_l$, also provides a
measure of $\theta.$
\begin{equation}
\label{idiff}
I_d=
-\frac V 2 \sin \phi \ \mathrm {erfi}
\left(\sqrt{-\log V}\right)
=-\frac V 2 \sin \phi \ \mathrm {erfi}\left(
kw\theta/\sqrt{8}\right)
\end{equation}
$I_s$ (Equation~\ref{Isvis}) and 
$I_d$ are relatively shifted by $\pi/2$ with respect to $\phi.$  
Taylor expanding Equation~\ref{idiff} for small $\theta$:
\begin{equation}
I_d\approx -\frac 1 {\sqrt{8\pi}}k\theta w\sin\phi
\left[1-\frac 1 {12} (k\theta w)^2\right]
=  -\frac 1 {\sqrt{8\pi}}k\theta w\sin\phi
\left[1-\frac 2 3 \log V\right]
\end{equation}

The sensitivity of $I_d$ to $\theta$ is 
\begin{equation}
\label{deriv-d}
\frac{dI_d}{d\theta}=-\frac 1 {\sqrt{8\pi}}kw\sin\phi
\left[1-\sqrt{\frac {\pi} {8}}
Vkw\theta  \ \mathrm {erfi}[
kw\theta/\sqrt{8}]\right].
\end{equation}
Equation~\ref{deriv-d} has the Taylor expansion
\begin{equation}
\label{deriv-dt}
\frac{dI_d}{d\theta}\approx-\frac 1 {\sqrt{8\pi}}kw\sin\phi
\left[1-\left(\frac{kw\theta}{2}\right)^2\right]
\approx -\frac 1 {\sqrt{8\pi}}kw\sin\phi
\left[2V-1\right].
\end{equation}

\subsection{Requirements on static alignment $\theta$ or alignment
sensing}
If the static misalignment $\theta$ were sufficiently small, the noise due
to alignment fluctuations $\tilde{\theta}(f)$ could be negligible, and would need
not be monitored.  Since the error in position readout associated with
offset $\phi$ from fringe center is $dz_{\phi} =  d\phi/(2k)$,
\begin{equation}
	\label{z0-err}
	\frac{dz_{\phi}}{d\theta}=\frac 1 {2k}\frac{d\phi}{d\theta}
=\frac{1}{2k}\frac{dI_s/d\theta}{dI_s/d\phi}
=	\frac{dz_{\phi}}{d\theta}=\frac 1 8 \frac {kw^2\theta}{\tan\phi}.
\end{equation}
Let $z_0$ be the operating point deviation from fringe center
$\phi=\pi/2.$  Then
$dz_{\phi}/d\theta=kw^2\theta\tan(2 k z_0)/8,$
implying the root Power Spectral Density
(RPSD) in displacement noise $\tilde{z}(f)$ for angle-noise
$\tilde{\theta}(f)$, fringe offset $z_0$, and alignment off-normal by
$\theta$ is
\begin{equation}
\label{z-fringe}
\tilde{z}_{\phi}(f) =\frac 1 8 \tilde{\theta}(f)\theta  kw^2\tan(2kz_0).
\end{equation}
For example, with $w=0.5$\,mm, 
$\tilde{\theta}(f)=1\,\mu$rad/$\sqrt{\rm Hz}$, static misalignment
$\theta = 100\,\mu$rad, and mass position offset $z_0=42$\,nm, the noise
contribution is $\tilde{z}(f)=10\,\mathrm{pm/\sqrt{Hz}}.$

While this error vanishes at the half-intensity point $z_0=0,$  two
other additive error terms are independent of $z_0.$  If the
interferometer beam strikes the test mass off-center by a distance
$h,$ then there is a ``sine error'' 
$dz_h/d\theta=(d/d\theta)(h\sin\theta)\approx h.$
There is also a ``cosine error'' associated with the combination of
wavefronts~\cite{logan}.  If the path from the test mass to the
beam-splitter is $p,$ the cosine error is
$dz_p/d\theta=(d/d\theta)(p\cos\theta)\approx -p\theta.$
Combining these terms with Equation~\ref{z0-err}, the
sensitivity of position readout $z$ to angle fluctuation $\theta$ is
\begin{equation}
	\frac{dz}{d\theta}\approx h+\theta\left[
	\frac{kw^2}{8}\tan(2kz_0)-p\right].
\end{equation}
Note that the first and last terms can be cancelled by aligning 
$\theta$ or $h$ to satisfy $h=p\theta.$  For
comparison,  the response of a heterodyne interferometer such as that
used in the LISA Test Package (LTP)
ESA counterpart of ST7 ~\cite{LTP} does not depend
on $z_0,$ and its $\theta$-sensitivity is simply $dz/dh \approx
h-p\theta.$  Eliminating $\theta$-sensitivity on ST7 requires either
holding $z_0$ fixed, or measuring $z_0$ (inferred from $I_s$)
and $\theta$ (inferred from $I_d$, Equation~\ref{idiff}).

\section{Experimental Results}
\begin{figure}
\begin{center}
\includegraphics[draft,bb=0 0 200 200]{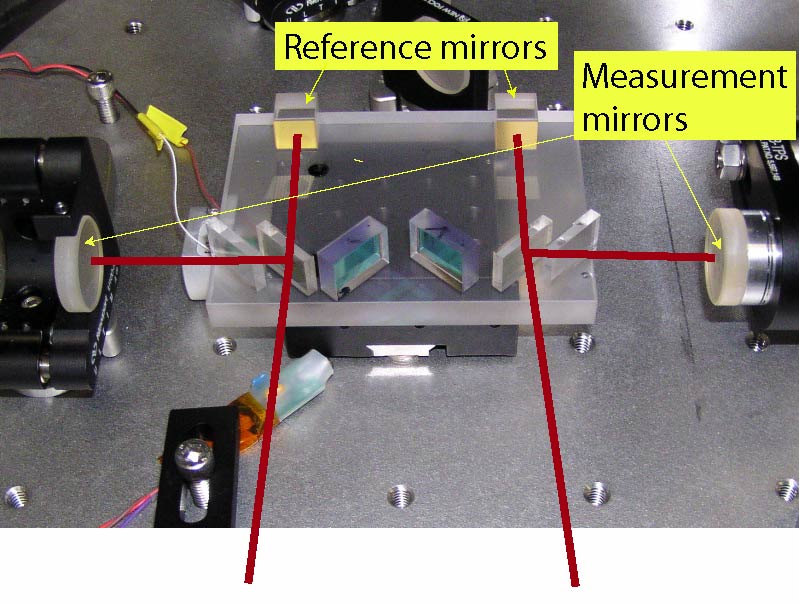}
\caption{Photograph of laboratory test interferometer.\label{photo}}
\end{center}
\end{figure}

Measurements were conducted on a laboratory test apparatus with
dimensions similar to those of ST7---see Figure~\ref{photo}.  The
Reference mirrors and beam-splitters, made from
ULE~\cite{ULE}  glass,  are optically contacted to a ULE plate.  
The plate is mounted on a
stage that can be moved in $z$ direction by a PZT actuator.
Measurement mirrors in separate, conventional mounts stand in for
the test masses.   One of the measurement mirrors is mounted on a
commercial~\cite{PI}  3-piezo actuator to allow $\theta$ angle actuation.
The spacing between the measurement mirrors is determined by a
stainless steel mounting plate, and the apparatus is in an evacuated
chamber.

\begin{figure}
\begin{center}
\includegraphics[scale=\figscale, angle=-90]{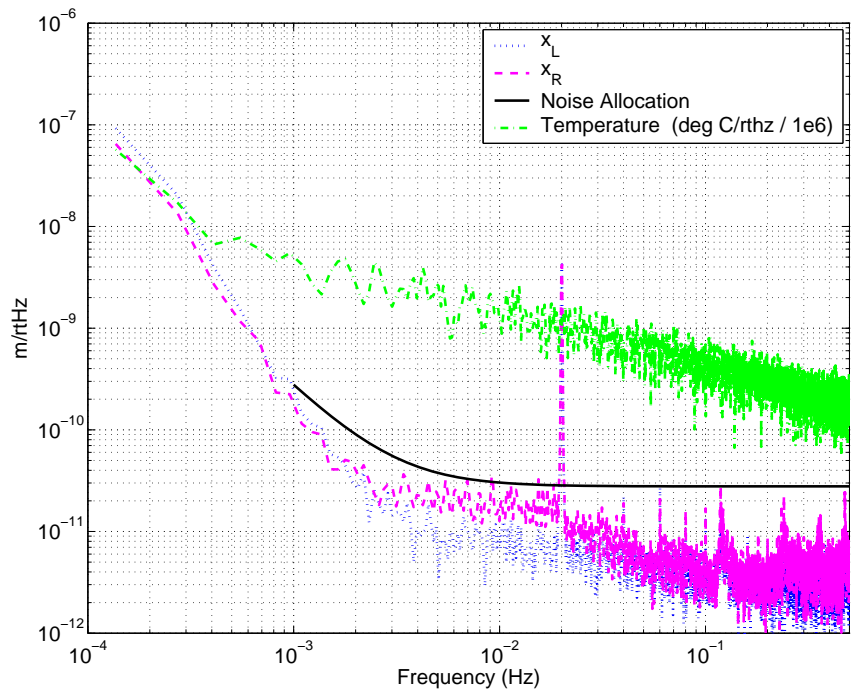}
\caption{Noise spectra from the laboratory test
interferometers, $x_L$ and $x_R$ measuring the left- and right-handed
paths.\label{spectra}}
\end{center}
\end{figure}

Noise performance of the interferometers is shown in
Figure~\ref{spectra}.  These data were collected six hours after the
chamber was evacuated, to allow the temperature to stabilize.  The
Noise Allocation curve is the error budget allocation to the RPSD
of $\tilde{z}(f).$  The noise levels for both the Left and Right 
interferometers are below
this allocation level, including $<30\,\mathrm{pm/\sqrt{Hz}}$ at 10\,mHz.  
The peak at 20\,mHz is due to intentional
sinusoidal modulation of $z,$ to verify calibration and to measure the
difference in the two interferometers' responses.  The calibration peaks
match to 0.5\%.  Also shown is the RPSD of the temperature in the
vacuum chamber.  Time-domain correlations are observed between the
interferometer output and the temperature for times on the order of
$10^4$\,sec and longer (not shown).

\begin{figure}
\begin{center}
\includegraphics[scale=\figscale,angle=-90]{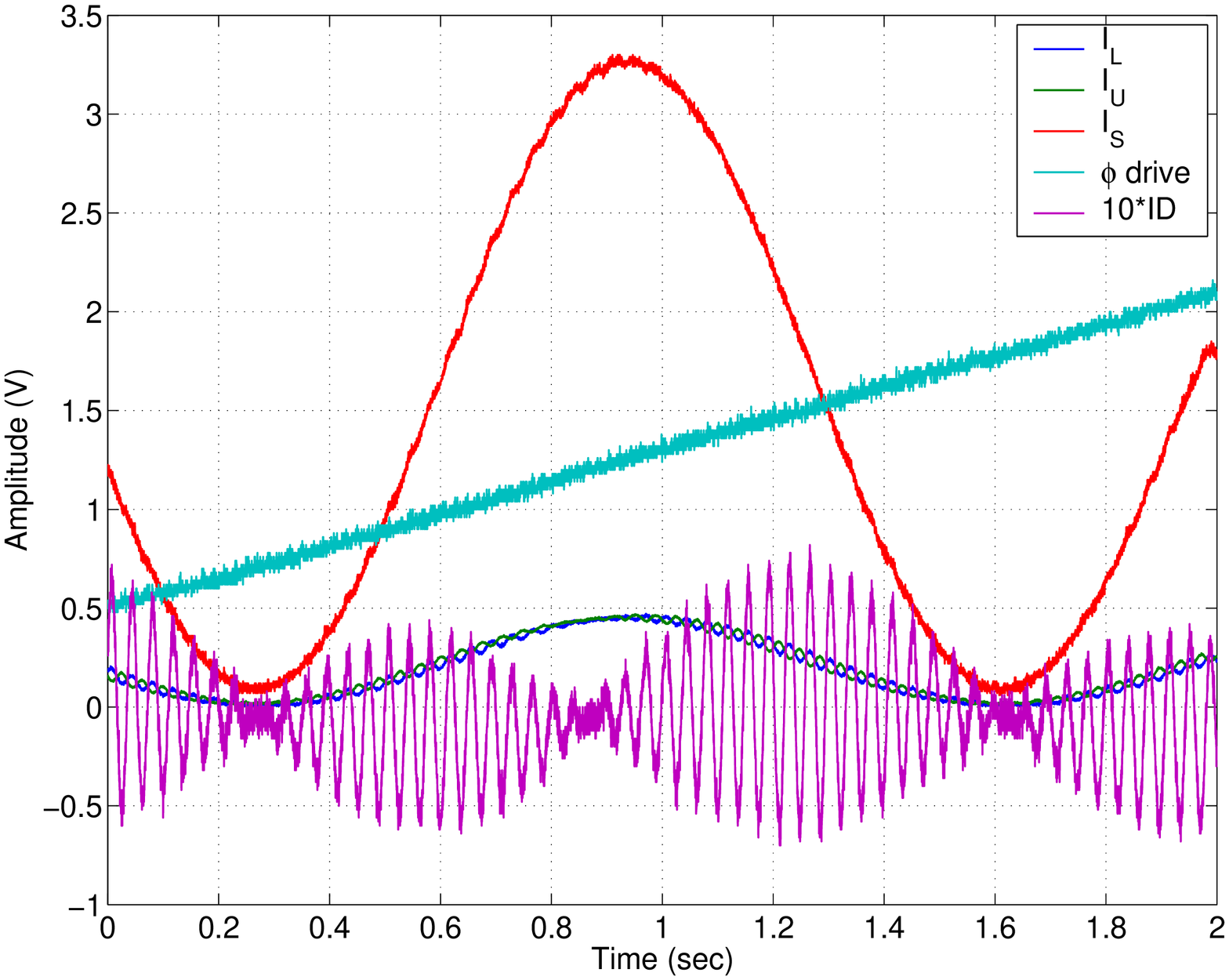}
\caption{Intensity measurements with $V\approx 1$ alignment, $z$ moved
at constant velocity, and  angular modulation $d\theta=
70\,\mu$rad p-p.  The $I_S$ trace is derived from a separate single-element
photoreceiver.
\label{algnd}}
\end{center}
\end{figure}

For the data shown in 
Figure~\ref{algnd},  the interferometer was aligned for best visibility
(minimum of $I_s$ nearly zero, $V\approx 1$), and
the optical bench was moved at constant velocity in the $z$ direction
while $\theta$ was modulated sinusoidally with  
amplitude $70\,\mu$rad p-p.
Ignoring the fast
$\theta$ response, the small phase shift
$\Delta \phi$ between $I_u$ and $I_l$ corresponds to $V\approx 1$,
in accordance with Equation~\ref{delta-phi}.

\begin{figure}
\begin{center}
\includegraphics[scale=\figscale,angle=-90]{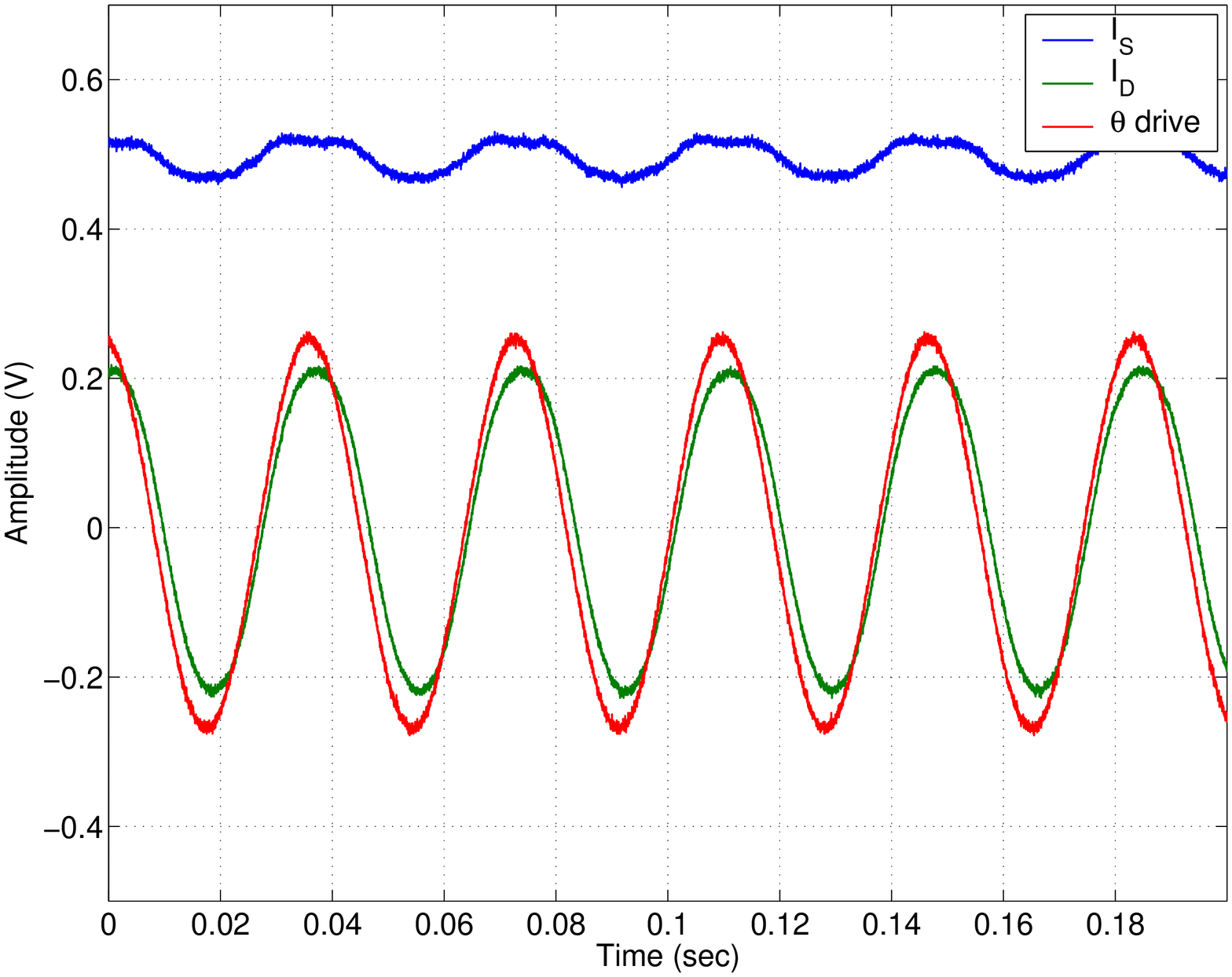}
\caption{Longitudinal position $z$ adjusted near the nominal half-intensity
point, $\phi=\pi/2.$ Transverse position $h$ adjusted for minimum
fluctuation in $I_s.$ $d\theta = 350\,\mu$rad p-p.
\label{ramp-mid}}
\end{center}
\end{figure}

Also in qualitative agreement with expectation is the response of the
difference signal to modulation,  Equation~\ref{deriv-d}.
At the half-intensity point, $\phi=\pi/2,$  $dI_d/d\theta$ is expected
to be maximum, corresponding to the maximum response to $\theta$
modulation in the experimental results.
The observed sensitivity of the sum signal to
$\theta$, $dI_s/d\theta$ also agrees qualitatively with the 
predicted $\phi$-dependence.
Equation~\ref{intens-sum-deriv} says that $dI_s/d\theta$ should be
maximum at the half-fringe, and the observed maxima are near
the half-fringe points.  

Figure~\ref{ramp-mid} shows that when the beam is centered ($h\approx
0$), the difference response $dI_d/d\theta$ is much larger than the sum
response $dI_s/d\theta$.  That is, under nominal operating conditions,
the diagnostic readout $dI_d$ of the error source $d\theta$ is larger
than its contribution to error in the signal, $dI_s.$

This research was performed at the Jet Propulsion Laboratory, California
Institute of Technology, under a contract to the National Aeronautics
and Space Administration.


\section*{References}
\begin{thebibliography}{99}
	\bibitem{PPA2} ``Laser Interferometer Space Antenna:  Pre-Phase
		A Report,'' 2nd ed, July 1998, available for public
		download from
		\url{ftp://ftp.ipp-garching.mpg.de/pub/grav/lisa/ppa2/ppa2.09.pdf}.
	\bibitem{ULE} A product of Corning Incorporated, Semiconductor
		Materials Business, Canton, New York,
		\url{http://www.corning.com/semiconductormaterials}
	\bibitem{logan} Logan J. et. al. 2002 {\it Applied Optics} {\bf
		41} 21
	\bibitem{PI} A product of Physic Instrumente (PI) GmbH \& Co.
		KG, Palmbach, Germany, \url{http://www.pi.ws}
	\bibitem{LTP} Vitale S., ``The SMART-2 LISA Test Flight,'' and
		Heinzel, G. ``The SMART-2 LTP IFO and Phasemeter,''
		these Proceedings.
\end{thebibliography}
\end{document}